\documentstyle[prl, aps, epsf, color, multicol]{revtex}
\begin{document}
\draft
\preprint{SBT2-JoGu20000307}

\title{
Alternative Buffer-Layers for the Growth of SrBi$_2$Ta$_2$O$_9$ on Silicon
	}
\author{
J. Schumacher, J. C. Mart\'\i nez, F. Martin, M. Maier, H. Adrian
	}
\address{
	Johannes Gutenberg - University of Mainz; Institute of Physics; 
	55099 Mainz; Germany
	}
\author{
R. Raiteri, H.J. Butt
	}
\address{
	Johannes Gutenberg - University of Mainz; Institute of Physical Chemistry; 
	55099 Mainz; Germany
	}
\date{March 7, 2000}
\maketitle

\begin{abstract}
In this work we investigate the influence of the use of YSZ and CeO$_2$/YSZ as insulators for Metal-
Ferroelectric-Insulator-Semiconductor (MFIS) structures made with SrBi$_2$Ta$_2$O$_9$ (SBT). We show 
that by using YSZ only the a-axis oriented Pyrochlore phase could be obtained. On the other hand the 
use of a CeO$_2$/YSZ double-buffer layer gave a c-axis oriented SBT with no amorphous SiO$_2$ inter-
diffusion layer. The characteristics of MFIS diodes were greatly improved by the use of the 
double buffer. Using the same deposition conditions the memory window could be increased from 
0.3 V to 0.9 V. From the piezoelectric response, nano-meter scale ferroelectric domains could be 
clearly identified in SBT thin films.
\end{abstract}


\begin{multicols}{2}
\narrowtext

\section{Introduction}
With the fast development of the semiconductor industry, there is nowadays a growing interest in the 
investigation of novel functional layers on silicon. In particular Perovskite materials have been 
intensively studied during the last years. That is mainly because of the large number of different 
physical properties observed in this class of materials. For example, we can find among perovskites, 
superconducting (YBa$_2$Cu$_3$O$_7$), ferromagnetic (La$_{0.67}$Ca$_{0.33}$MnO$_3$) and ferroelectric compounds 
(Pb(Zr$_{1-x}$Ti$_x$)O$_3$).

Due to the high reactivity of silicon with oxygen, the deposition of high quality perovskites on similar 
substrates is no trivial task. However during the last years, the possibility of growing crystalline 
perovskites on Si by Pulsed Laser Deposition (PLD) was widely demonstrated \cite{1}. Although 
this technology is not yet comfortable for wide area deposition, complex heterostructures can be 
prepared.

More recently, because of a growing interest on the development of non-volatile memory cells, 
several groups concentrated their activities on the deposition of SrBi$_2$Ta$_2$O$_9$ thin films (SBT) on 
silicon by Pulsed Laser Deposition \cite{2}. A large number of buffer layers like Y$_2$O$_3$ \cite{3}, 
SrTiO$_3$, MgO \cite{4}, Si$_3$N$_4$ \cite{5} and Al$_2$O$_3$ \cite{6,7} have been investigated. However the 
formation of an amorphous SiO$_2$ inter-diffusion layer can hardly be controlled by the use of these 
buffers. 

In this work, we investigate the influence of (Y$_2$O$_2$)$_x$(ZrO$_2$)$_{1-x}$ (YSZ) and CeO$_2$/YSZ buffer layers 
on the structure, surface morphology and ferroelectric properties of SrBi$_2$Ta$_2$O$_9$ thin films. The YSZ 
layer has been used in order to avoid the formation of SiO$_2$. 

First a description of the employed deposition method is described and the structural properties of the 
different layers are discussed. The characteristic surface and ferroelectric domain morphologies are 
studied by using Atomic Force Microscopy (AFM). Finally we compare the Capacitance versus 
bias Voltage (CV) characteristics of Metal-Ferroelectric-Semiconductor capacitors for SrBi$_2$Ta$_2$O$_9$ 
layers deposited with and without buffers.

\section{ PLD deposition with different buffer layers }

One of the challenging problems in depositing crystalline oxide layers on Si is to avoid the 
spontaneous formations of SiO$_2$ at the interface. Thanks to the discovery of High Temperature 
Superconductors (HTS), a large effort was done in order to get an effective solution. This was 
mainly because the superconducting properties of HTS can only be obtained through a reasonable 
crystalline quality. Here the solution was to use a YSZ buffer layer (Y$_2$O$_2$)$_x$(ZrO$_2$)$_{1-x}$ in order to 
avoid the formation of SiO$_2$. Later additional layers of BaZrO$_3$ or CeO$_2$ have been used in order to 
avoid inter-diffusion and increase the crystalline quality. From the different studies it became clear 
that during the deposition of the YSZ layer, Zr reacts at high temperatures with SiO$_2$ giving ZrO$_2$ and 
SiO \cite{1}. The latter can be easily removed by natural sublimation which occurs at low pressures 
(below 10$^{-6}$ mbar) and temperatures above 800¡C. The best YSZ layers are usually obtained at 
pressures below 10$^{-5}$ mbar \cite{fork}. The YSZ layer has to be deposited at low pressures 
while the other layers need usually pressures larger than 0.3 mbar. An in-situ deposition of the 
different layers, at large deposition rates, can be easily achieved by Pulsed Laser Deposition (PLD). 
However an ex-situ deposition of perovskites on top of pre-buffered substrates is also possible \cite{8}.

Our PLD chamber was specially designed for depositing complex heterostructures. The system 
consists of a carousel-type holder, which allows the in-situ deposition of up to six layers. The 
deposition chamber has a base pressure of about 10$^{-7}$ mbar and maximum deposition temperature of 
about 1000$^\circ$C. A self designed computer controlled system, allows the scanning of the laser plum 
over an area of about 2$\times$2 cm$^2$.

The silicon substrates were cleaned with different steps. In order to remove organic contamination, 
the substrates were cleaned with a solution of H$_2$SO$_4$ and H$_2$O$_2$. The native SiO$_2$ layer was etched 
by a standard HF solution.

\begin{figure}[t]
	\centering
	\epsfxsize = 8 cm
	\epsfbox{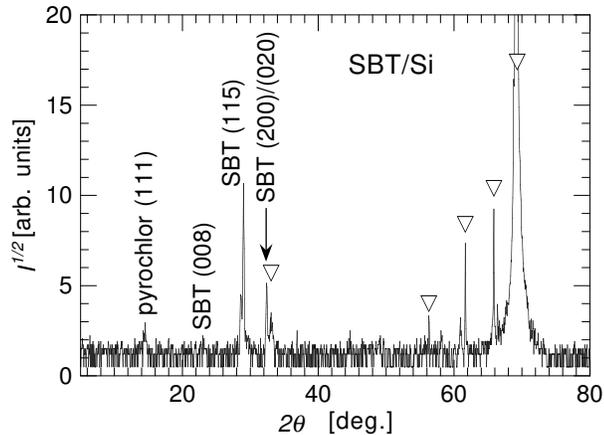}
	\vspace{5 mm}
	\caption{
X-ray diffraction in Bragg-Brentano geometry of a SBT film deposited directly on 
silicon. The triangles mark the reflections corresponding to the substrate.
\label{fig1}}
\end{figure}

The best SBT films deposited on Si were obtained at substrate temperatures of 800$^\circ$C and O$_2$ 
deposition pressures of 0.8 mbar. Only at lower pressures we observed the formation of the 
Pyrochlore SBT phase (p-SBT). Fig.\ \ref{fig1} shows a $\theta$-$2\theta$ scan of a SBT film deposited on Si. It can be 
observed that a small amount of the p-SBT phase is still present. From low angle x-ray diffraction we 
detected a 4 nm thick layer which was attributed to a SiO$_2$ inter-diffusion layer. Because of the 
amorphous SiO$_2$, the SBT layer did not grow with a defined orientation. The intensities observed in 
Fig.\ \ref{fig1} are consistent with the expected values for a non-oriented powder.

%
%

In order to get rid of the 4 nm SiO$_2$, a 40 nm thick YSZ buffer layer was deposited at typically 850$^\circ$C 
and 10$^{-5}$ mbar. Under these conditions the YSZ layer is c-axis oriented with rocking curves in the 
order of 1.2 degrees. The presence of the (111) reflection shows that a small amount of randomly 
oriented crystallites is still present. However, from x-ray low-angle diffraction experiments, we could 
not detect any SiO$_2$ layer. 

%
%
\begin{figure}[t]
	\centering
	\epsfxsize = 8 cm
	\epsfbox{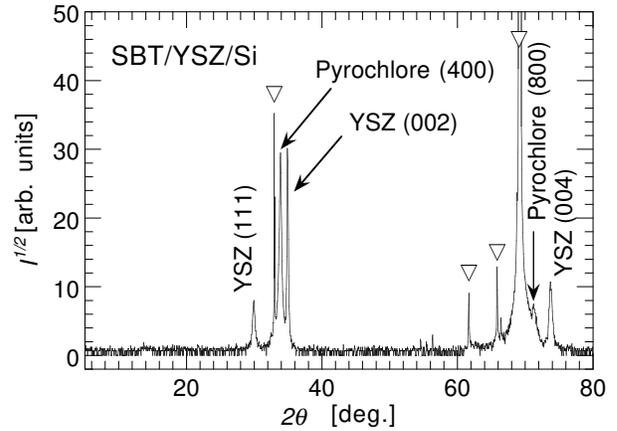}
	\vspace{5 mm}
	\caption{
X-ray diffraction of a SBT film deposited in a YSZ buffer layer. Despite the fact that 
the deposition parameters were identical as those in Fig.\ \ref{fig1}, we got mainly the Pyrochlore 
phase.
\label{fig2}}
\end{figure}
%
%
\begin{figure}[t]
	\centering
	\epsfxsize = 8 cm
	\epsfbox{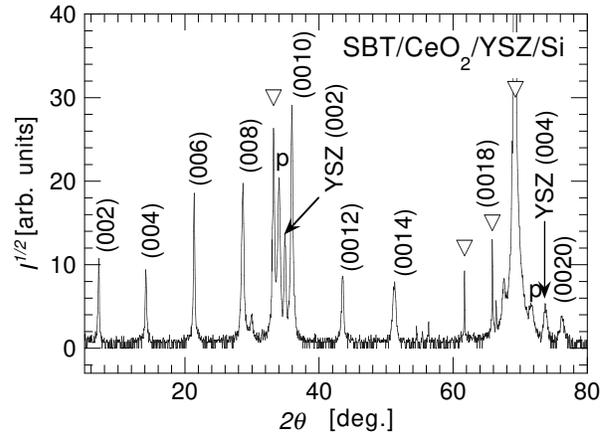}
	\vspace{5 mm}
	\caption{
SBT deposited on top of a CeO2/YSZ buffer shows c-axis growth with a rocking curve 
of 1.2$^\circ$ for the (006) reflection.
\label{fig3}}
\end{figure}

\begin{figure}[t]
	\centering
	\epsfxsize = 8 cm
	\epsfbox{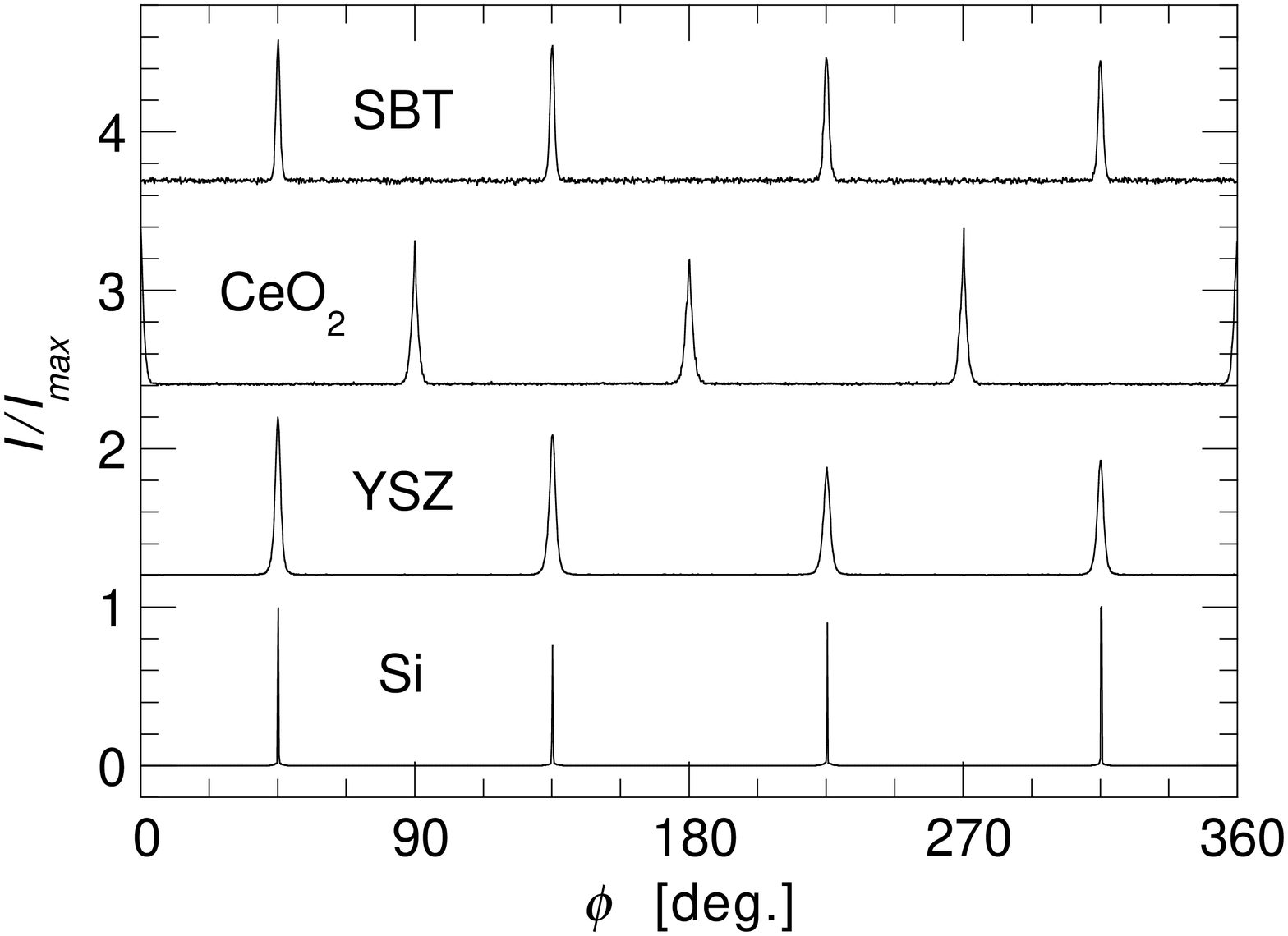}
	\vspace{5 mm}
	\caption{
$\Phi$-scans show that SBT films grows cube to cube with respect to the silicon substrate. 
For SBT, YSZ and Si, we used the (111) reflexes. For CeO$_2$, the (024) 
reflexes were taken into account because of the small lattice mismatch with respect to Si.
\label{fig4}}
\end{figure}

As we can see in Fig.\ \ref{fig2}, for the same deposition parameters for SBT as in Fig.\ \ref{fig1}, we could only 
grow an a-axis oriented p-SBT phase with a rocking curve of 1.4 degrees for the (400) reflex. No 
temperature/pressure-window could be found in order to get the desired SBT phase. This result is consistent
with the work of Ishikawa et al. done on YSZ substrates \cite{ishikawa}.
As it is shown in Fig.\ \ref{fig3}, by introducing a 20 nm thick CeO$_2$ layer between SBT and YSZ we 
succeeded in obtaining a fully c-axis oriented SBT film with rocking curve of 1.2$^\circ$ for the (006) 
reflection. One can notice that only a small amount of p-SBT phase is still present. The in-plane 
orientation of the films was checked by $\Phi$-scans. It can be seen that, indeed, the SBT films grow 
cube-to-cube with respect to the silicon substrate (Fig.\ \ref{fig4}). The biggest difficulty, in this case, was the 
small lattice mismatch between CeO$_2$ and Si, which obliged us to measure the (024) family of 
reflections for CeO$_2$.   

%
%

%
%
\begin{figure}[t]
	\centering
	\epsfxsize = 5 cm
	\epsfbox{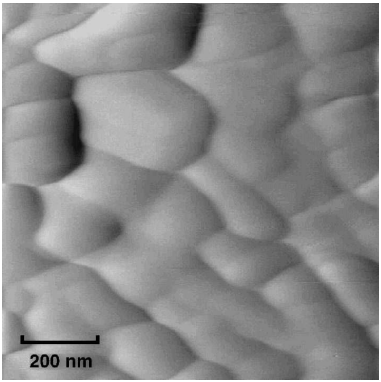}
	\vspace{5 mm}
	\caption{
 AFM contact topography of a 300 nm 
 thick SBT thin film deposited on Si. Vertical scale 
 is 20 nm
\label{fig5}}
\end{figure}
%
%
\begin{figure}[t]
	\centering
	\epsfxsize = 5 cm
	\epsfbox{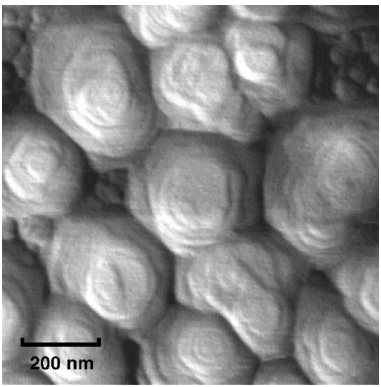}
	\vspace{5 mm}
	\caption{
 AFM contact topography of a 
 SBT/CeO2/YSZ/Si heterostructure. The terraces 
 correspond to unit-cell steps along the c-axis. 
 Vertical scale is 20 nm
\label{fig6}}
\end{figure}

\section{ Characterization by Atomic Force Microscopy }

We studied the topography of the different ferroelectric films with a commercial Atomic Force 
Microscope (AFM) \cite{9} (Multimode, Digital Instruments, Santa Barbara, CA).  The surface 
roughness was directly calculated from the topography images obtained in contact mode. For the 
SBT/Si layers we obtained an average roughness value of 5.6 nm rms (Fig.\ \ref{fig5}). An AFM image of the 
c-axis oriented SBT/CeO$_2$/YSZ/Si is shown in Fig.6. In this case the average roughness is 4.4 nm 
rms. The terraces in Fig.6 correspond to the c-axis unit cell of SBT, while the smaller grains 
observed at the top-right side correspond to the p-SBT phase, which could be observed in the x-ray 
data too. 

%
%
\begin{figure}[t]
	\centering
	\epsfxsize = 4 cm
	\epsfbox{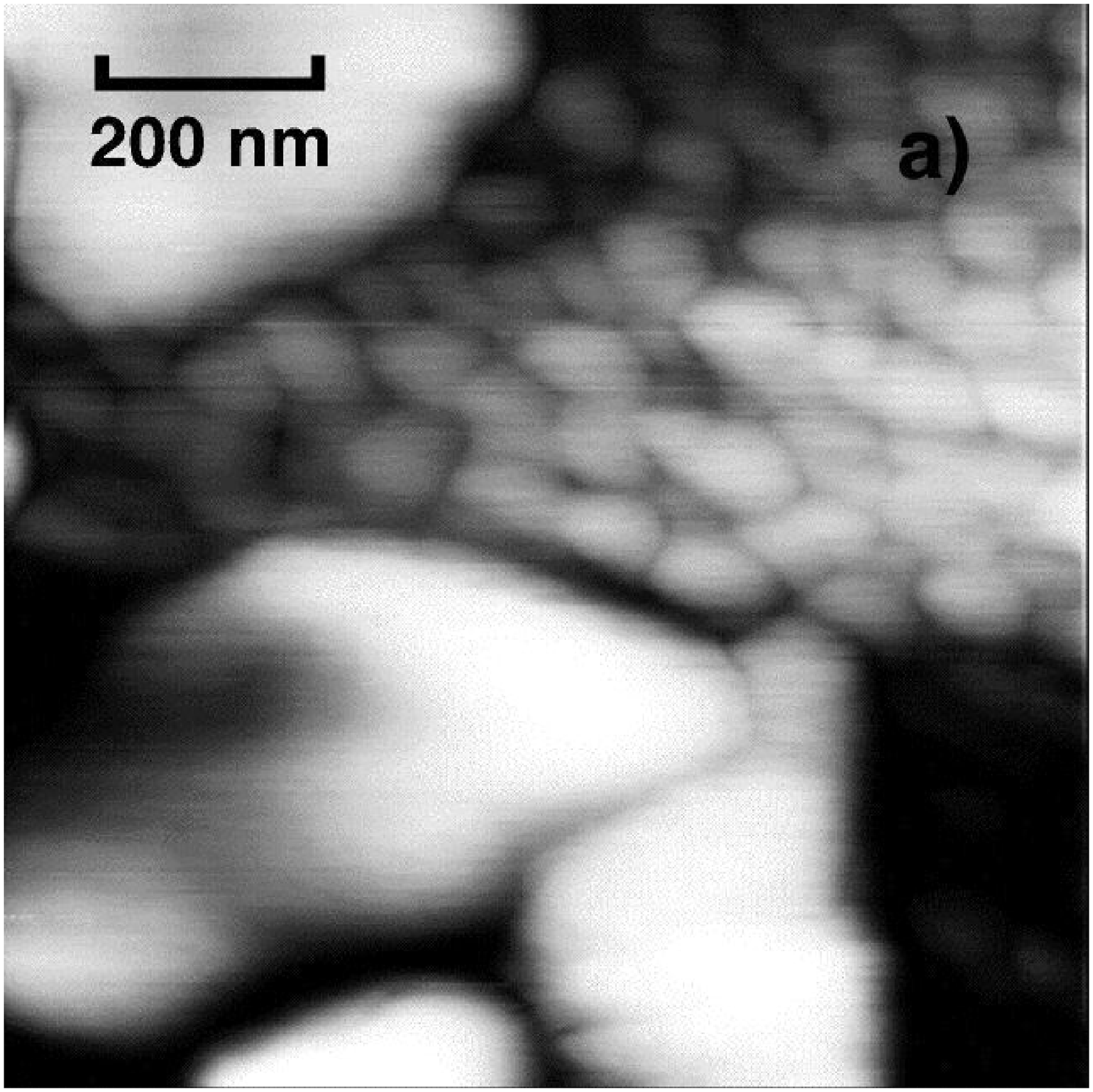}
	\centering
	\epsfxsize = 4 cm
	\epsfbox{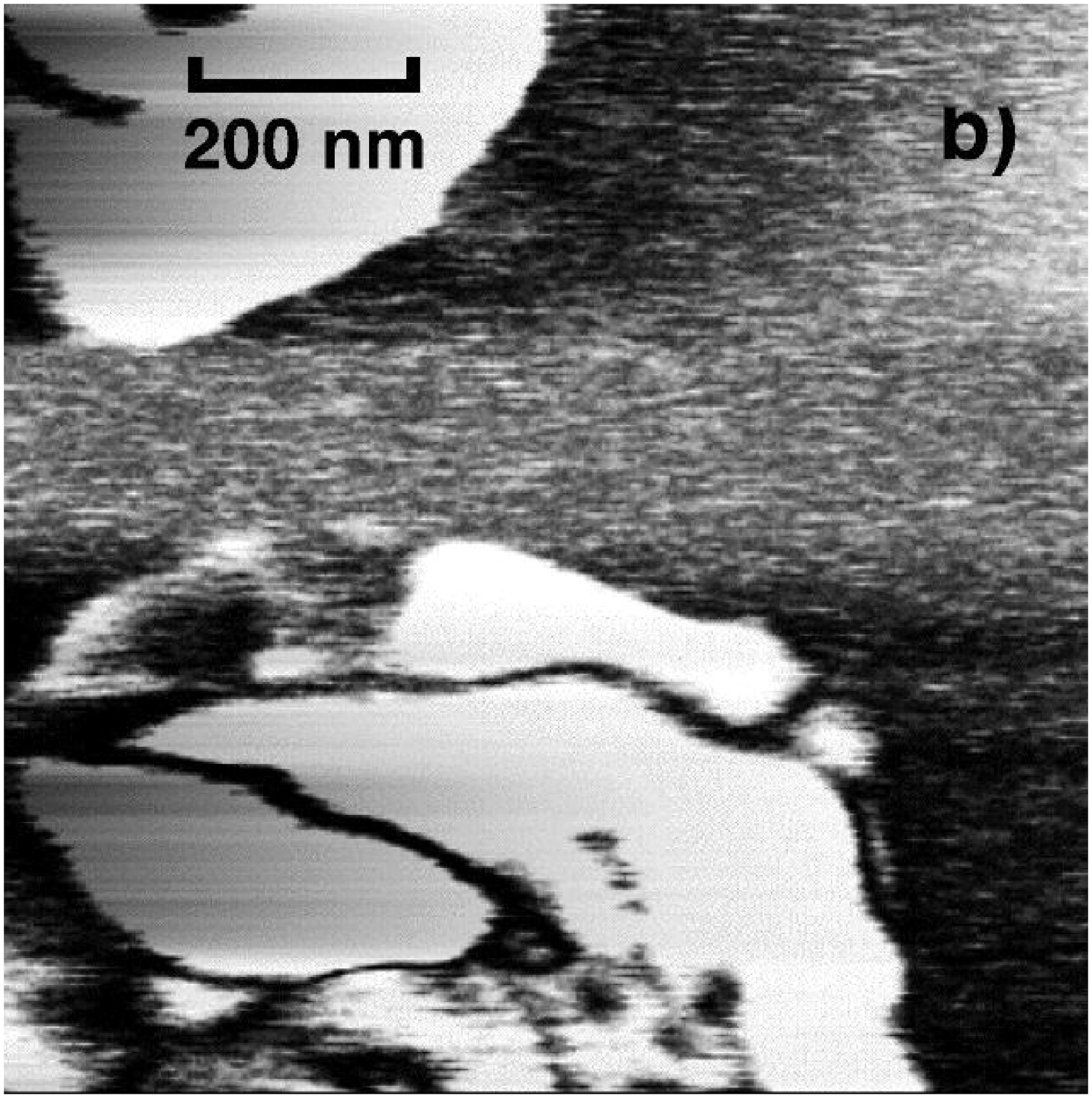}
	\centering
	\epsfxsize = 4 cm
	\epsfbox{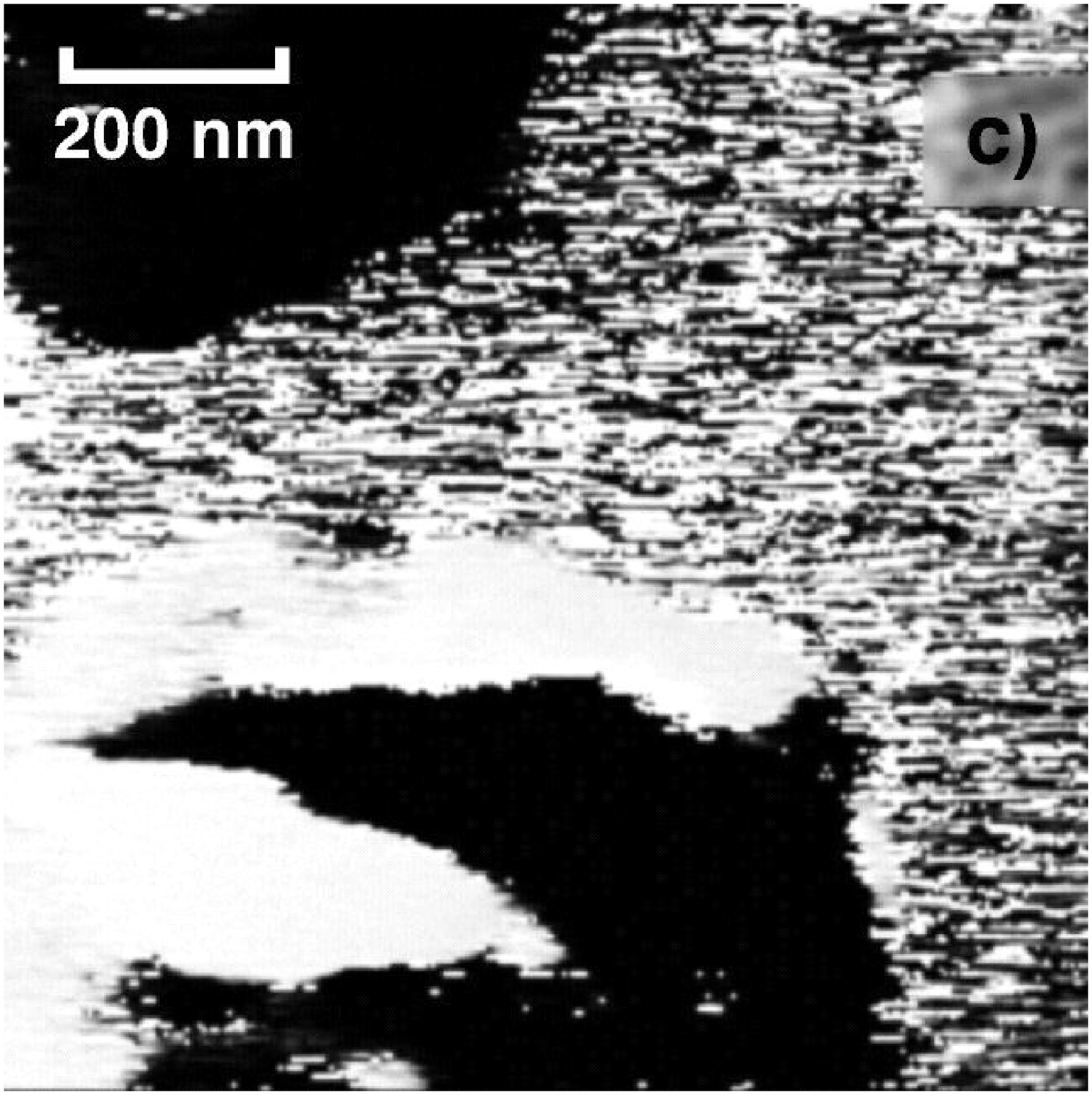}
	\vspace{5 mm}
	\caption{
 AFM pictures of the same area of a SBT film grown on silicon. In this case, the film 
 was grown at lower pressures in order to enhance the Pyrochlore phase, which corresponds 
 to the 50 nm-wide grains in the topography image a) (z-range is 50 nm). Images b) and c) 
 show, respectively, the amplitude and phase shift of the piezoelectric-response to an AC 
 voltage (15 Vpp, 10 kHz). In image b) lighter colour means a higher piezoelectric response, 
 full scale is 0.05 nm. In image c) contrast is given by relative phase shift, full scale is 180$^\circ$.
\label{fig7}}
\end{figure}

In order to characterise the ferroelectric domains, an alternating voltage was applied between the 
conductive AFM tip and the back-side of the sample substrate during imaging, as already described in 
details by Martin et al. and Auciello et al. \cite{9}. The piezoelectric response to the 
applied AC voltage is probed by sub-nanometer oscillations of the tip, superimposed to the static 
deflection kept by the AFM feedback loop. Phase shift and amplitude of the cantilever oscillation can 
be detected by a lock-in amplifier and recorded simultaneously with the sample topography.

Fig.\ \ref{fig7} shows the topography, amplitude and phase shift of a 300 nm thick SBT film. This film was 
deposited at lower O$_2$ pressure (0.3 mbar) so that it is possible to observe as well the response of the 
50 nm wide p-SBT grains, which appear in Fig.\ \ref{fig7}, between two large SBT grains. From the phase 
signal, it is possible to identify ferroelectric domains with opposite dipole orientation inside the SBT 
grains. In the amplitude image, we distinguish the domain walls (darker regions) between the 
ferroelectric domains (bright regions). The p-SBT phase shows no evident piezoelectric response.

Similar experiments done in c-axis oriented SBT films did not show features that we could directly 
correlate to domain structures, although, as it will be shown below, the films still showed a 
ferroelectric behaviour.

\section{Metal-Ferroelectric-Insulator-Semiconductor structures}

The CV curves were found to be weakely dependent on the AC-amplitude for a range between 50 and 100 mV. 
The gold electrodes had a typical area of 0.5$\times$0.5 mm$^2$.

\begin{figure}[t]
	\centering
	\epsfxsize = 8 cm
	\epsfbox{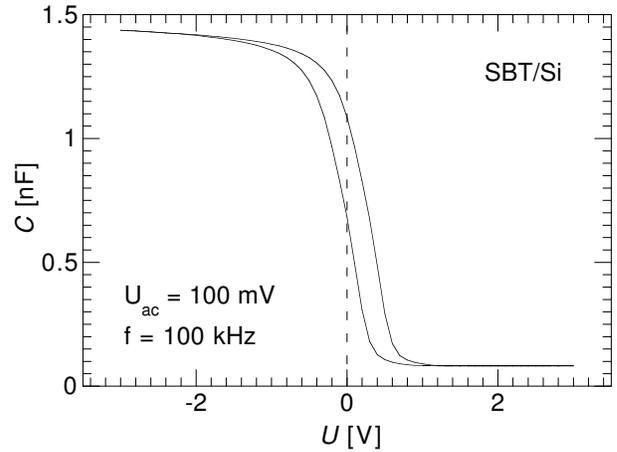}
	\vspace{5 mm}
	\caption{
Capacitance versus Voltage diagram for a metal-ferroelectric-semiconductor 
structure (MFS). The measurement was done at 100 kHz with AC-amplitude of 100 mV. The 
memory-window is 0.3 V wide
\label{fig8}}
\end{figure}

Fig.\ \ref{fig8} shows the capacitance versus bias Voltage U of a Au/SBT/Si heterostructure with a 300 nm 
thick SBT layer.  The behaviour is very similar to what is expected for a Metal-Ferroelectric-
Semiconductor diode measured at high frequency. Additionaly the hysteresis induced by the coercive field of SBT 
ca be observed \cite{10}. The ferroelectricity of our isotropic SBT-film generates a typical memory window of 
0.3 V.

%
%
%
%
\begin{figure}[t]
	\centering
	\epsfxsize = 8 cm
	\epsfbox{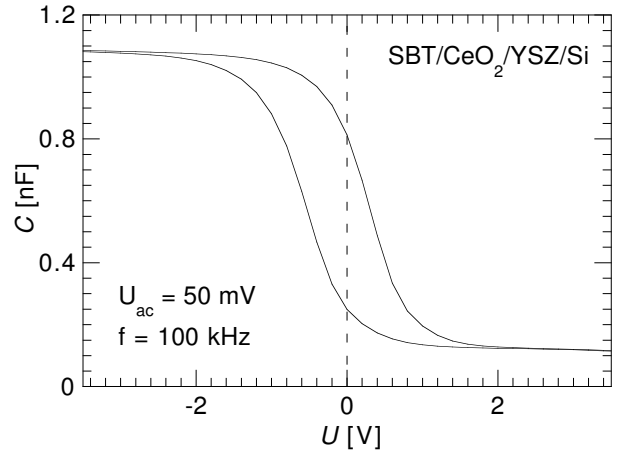}
	\vspace{5 mm}
	\caption{
Capacitance versus Voltage diagram for a Au/SBT/CeO2/YSZ/Si heterostructure. This 
measurement was done at 100 kHz with AC-amplitude of 50 mV. The memory-window is 0.9 
V wide.
\label{fig9}}
\end{figure}

Fig.\ \ref{fig9} shows a CV-characteristic for a Au/SBT/CeO$_2$/YSZ/Si heterostructure. The thicknesses 
were 50, 20, and 40 nm for SBT, CeO$_2$, and YSZ respectively. 
In this case, the use of the buffer layers increased the memory window to 0.9 V. The remanent capacitance can be 
switched between 200 and 800 pF for voltages between $\pm$ 3 V. As it was observed by Han et al \cite{10} 
and predicted by Miller and McWhorter \cite{11}, the memory window is mainly related to the 
coercitive field and is very little influenced by the amplitude of the remanent polarization (for P$_r >$ 0.1 
µC/cm$^2$).

\section{Conclusion}

We have shown that by using CeO$_2$/YSZ buffers it is possible to obtain c-axis oriented SBT films 
with rocking curves of 1.2$^\circ$ for the (006) reflection. $\Phi$-Scans show that the films grow cube-to-cube 
with respect to the silicon substrate. In spite the fact that the easy direction for the polarization of SBT 
is known to lay along the (a,b)-plane, we observed, for c-axis SBT films, a memory window of 
MFIS structures of 0.9 V instead of the 0.3 V measured in Au/SBT/Si. The capacitance of the device could 
be switched by a factor 4 (from 200 to 800 pF) by applying  3 V. 

These improvements are mainly due to the better crystalline quality of the c-axis-oriented films and 
the absence of amorphous SiO$_2$ at the YSZ/Si interface.

From AFM measurements we have been able to identify terraces which correspond to the c-axis unit 
cell, and we could observe ferroelectric domains in non-oriented films. For the moment we could not 
identify a clear domain structure in c-axis oriented SBT films.

\section{Acknowledgement}

J.S would like to acknowledge support from the Undergraduate Materials Research Initiative from the 
Materials Research Society. J.C.M and R.R were financed by the European Union under contracts 
(ERBMBICT972217 and FMBICT972487). This work was partially supported by the Material 
Wissenschaftliches Forschung Zentrum of the University of Mainz.

\end{multicols}
\end{document}